\documentclass[aps,prl,twocolumn,superscriptaddress,groupedaddress]{revtex4}
\pdfoutput=1

\usepackage[dvipsnames]{xcolor}
\usepackage{graphicx}
\usepackage{natbib}
\usepackage{amsmath,empheq}
\usepackage{ulem}
\usepackage{dcolumn}
\usepackage{bm}

\newcommand{\la}{\left<}
\newcommand{\ra}{\right>}

\begin{document}

\preprint{APS/123-QED}
\title{Effective transport by 2D turbulence: \\
Vortex-gas theory vs. scale-invariant inverse cascade}

\author{Julie Meunier}
\author{Basile Gallet}
\affiliation{Université Paris-Saclay, CNRS, CEA, Service de Physique de l’Etat Condensé, 91191 Gif-sur-Yvette, France.}

\begin{abstract}

The scale-invariant inverse energy cascade is a hallmark of 2D turbulence, with its theoretical energy spectrum observed in both Direct Numerical Simulations (DNS) and laboratory experiments. Under this scale-invariance assumption, the effective diffusivity of a 2D turbulent flow is dimensionally controlled by the energy flux and the friction coefficient only. Surprisingly, however, we show that such scaling predictions are invalidated by numerical solutions of the 2D Navier-Stokes equation
forced at intermediate wavenumber and damped by weak linear or quadratic drag. We derive alternate scaling-laws for the effective diffusivity based on the emergence of intense, isolated vortices causing spatially inhomogeneous frictional dissipation localized within the small vortex cores. The predictions quantitatively match DNS data. This study points to a universal large-scale organization of 2D turbulent flows in physical space, bridging standard 2D Navier-Stokes turbulence with large-scale geophysical turbulence.
\end{abstract}

\keywords{Suggested keywords}

\maketitle


\textit{Introduction.—} Many turbulent flows are effectively described by two-dimensional (2D) equations as a result of small aspect ratio, global rotation or external magnetic field~\cite{sommeria1982and,davidson2013turbulence,deusebio2014dimensional,gallet2015exacta,gallet2015exactb,benavides2017critical,seshasayanan2018condensates,ALEXAKIS20181,van2019condensates,seshasayanan2020onset}, in situations ranging from large-scale flows in oceans and planetary atmospheres~\cite{salmon1998lectures,pedlosky2013geophysical,vallis2017atmospheric,miller2024gyre}
to magneto-hydrodynamics~\cite{driscoll1990experiments,driscoll2002vortex,sommeria1982and,sommeria1986experimental,gallet2012reversals,michel2016bifurcations} and active matter systems~\cite{martinez2021scaling,bardfalvy2024collective}. The phenomenology of 2D turbulence strongly departs from that of 3D turbulence. The conservation of both energy and enstrophy by 2D flows induces inverse energy transfers, from smaller to larger scales, together with forward enstrophy transfers, from larger to smaller scales. When energy is supplied at wavenumber $k_f$ and removed at large scales by some damping mechanism, the Kraichnan-Leith-Batchelor (KLB) scaling theory predicts an inverse energy cascade characterized by an energy spectrum proportional to $k^{-5/3}$, where $k$ denotes the wavenumber~\cite{kraichnan1967inertial}.

The $k^{-5/3}$ energy spectrum has been observed in both laboratory experiments and Direct Numerical Simulations (DNS)~\cite{sommeria1986experimental,boffetta2000inverse,chen2006physical,boffetta2010evidence,boffetta2012two,ALEXAKIS20181}, making the KLB theory standard textbook material~\cite{lesieur2008introduction,frisch1995turbulence}. Beyond the sole energy spectrum, however, one is often interested in the large-scale organization of the flow and in its effective transport properties as it acts on a much larger-scale tracer distribution: given a large-scale background gradient of some passive tracer, what is the mean scalar flux induced by the 2D turbulent flow, or equivalently, what is the effective diffusivity of the turbulent flow? On the one hand, there is a vast literature on inertial-range statistics of concentration fluctuations in passive scalar turbulence ~\cite{kraichnan1994anomalous,celani2001statistical,cardy2008non, shraiman2000scalar}. On the other hand, the effective diffusivity has been characterized only within the analytically tractable regime of laminar flows at moderate Péclet number~\cite{sivashinsky1985weak,rosenbluth1987effective,dubrulle1991eddy,biferale1995eddy}. By contrast, the effective diffusivity of a fully turbulent 2D flow has received far less attention, although it plays a central role in the context of oceanic and atmospheric flows (see {\it discussion} section). 

In the following we thus report a numerical and theoretical study on the effective diffusivity of 2D turbulent flows. 
The DNS flow is driven at a scale much smaller than the domain size. It is damped by a drag force, either linear or quadratic in velocity. 
Linear drag is motivated by the Hartmann friction of quasi-2D magnetohydrodynamic flows~\cite{sommeria1986experimental} and the Ekman friction of quasi-2D rapidly rotating flows~\cite{salmon1998lectures,pedlosky2013geophysical,vallis2017atmospheric}. To better model the turbulent drag force on the roughness elements of the Earth surface and Ocean floor, atmospheric and Ocean models often resort to a quadratic drag force instead, proportional to the local velocity squared~\cite{holton2013introduction,grianik,ChangHeld}.

The core assumptions of the KLB theory -- a scale-invariant inverse-energy cascade -- readily provide standard dimensional estimates \textit{à la Kolmogorov} for the effective diffusivity of the flow: the latter is dimensionally constrained by the energy flux and the drag coefficient arresting the cascade only, while being independent of the forcing scale $k_f^{-1}$~\cite{SMITH2002,grianik}. We refer to the resulting scaling-laws for the effective diffusivity as the scale-invariant inverse cascade (SIIC) scaling predictions.
Surprisingly, however, the present DNS data indicate that the effective diffusivity of the flow strongly departs from such SIIC dimensional estimates. We attribute the failure of the scale-invariance argument to the emergence of coherent vortices~\cite{borue1994inverse,danilov2001nonuniversal,scott2007nonrobustness,burgess2017vortex}: the vortices come with two characteristic scales, the inter-vortex distance and the vortex-core radius, thus breaking the scale-invariance of the flow.

Also suggestive of the importance of coherent vortices is the fact that a successful vortex-based scaling theory for freely evolving 2D turbulence has been around for decades \cite{mcwilliams1990vortices,Carnavaleetal}. Finally, Chang \& Held noticed that the KLB dimensional estimates are not satisfied by the quasi-2D `baroclinic' turbulence arising in models of large-scale ocean and atmospheric flows with quadratic drag~\cite{ChangHeld}. In this context, the crucial role of large-scale vortices~\cite{thompson2006scaling} is the starting point of a scaling theory put forward by Gallet \& Ferrari (\cite{GalletFerrari}, GF in the following), but a connection to standard 2D Navier-Stokes turbulence has so far remained elusive. The goal of this Letter is to establish this connection through the development of a quantitative vortex-based theory for the eddy diffusivity of 2D Navier-Stokes turbulence with drag. 

\textit{Forced 2D turbulence with drag.—} We consider a two-dimensional incompressible flow ${\bf u}=(u,v) = (-\psi_y, \psi_x)$ inside a doubly periodic domain $[0,2 \pi L]^2$. The flow is driven at small scale by the curl of a body-force, $f(x,y,t)$, peaked around a wavenumber $k_f$. 
The governing equation for the vorticity $\zeta = \Delta \psi$ reads
\begin{eqnarray}
\partial_t \zeta + J(\psi,\zeta) = -\mathcal{D}(\psi) + f(x,y,t) - \nu  \Delta^4 \zeta  \, , \label{eq:zeta}
\end{eqnarray}

Where $J(g, h) = g_x h_y - g_y h_x$ denotes the Jacobian operator and $\nu$ denotes the hyperviscosity coefficient. The drag term $\mathcal{D}(\psi)$ corresponds to (the curl of) either a linear or a quadratic drag force. Denoting as $\kappa$ (resp. $\mu$) the linear (resp. quadratic) drag coefficient, the drag term reads:
\begin{eqnarray}
    \mathcal{D}(\psi) = \left\{
    \begin{array}{ll}
        \kappa \Delta \psi & \mbox{linear drag} \\
        \mu \left[\left( |\nabla \psi| \psi_x \right)_x + \left( |\nabla \psi| \psi_y \right)_y \right]& \mbox{quadratic drag.}
    \end{array}
    \right.
   \nonumber
\end{eqnarray}

We consider two small-scale forcing protocols. Following the most widely adopted setup in the literature~\cite{SMITH2002,grianik,chen2006physical,boffetta2010evidence,boffetta2012two}, in a first suite of numerical runs $f(x,y,t)$ is white-noise-in-time isotropic forcing inside a narrow band of wavenumbers centered around $k_f$.
In a second suite of numerical runs $f(y)$ is steady `Kolmogorov' forcing proportional to $\cos(k_f y)$ \cite{burgess1999instability,rivera2000external,TsangYoung}. For the former forcing the mean energy injection rate per unit mass of fluid $\epsilon$ is a control parameter of the system, while for the latter forcing $\epsilon$ is an emergent parameter. Regardless, we use $\epsilon$ to characterize the strength of the forcing in both cases.

\begin{figure}[h!]
\includegraphics[width=8.5cm]{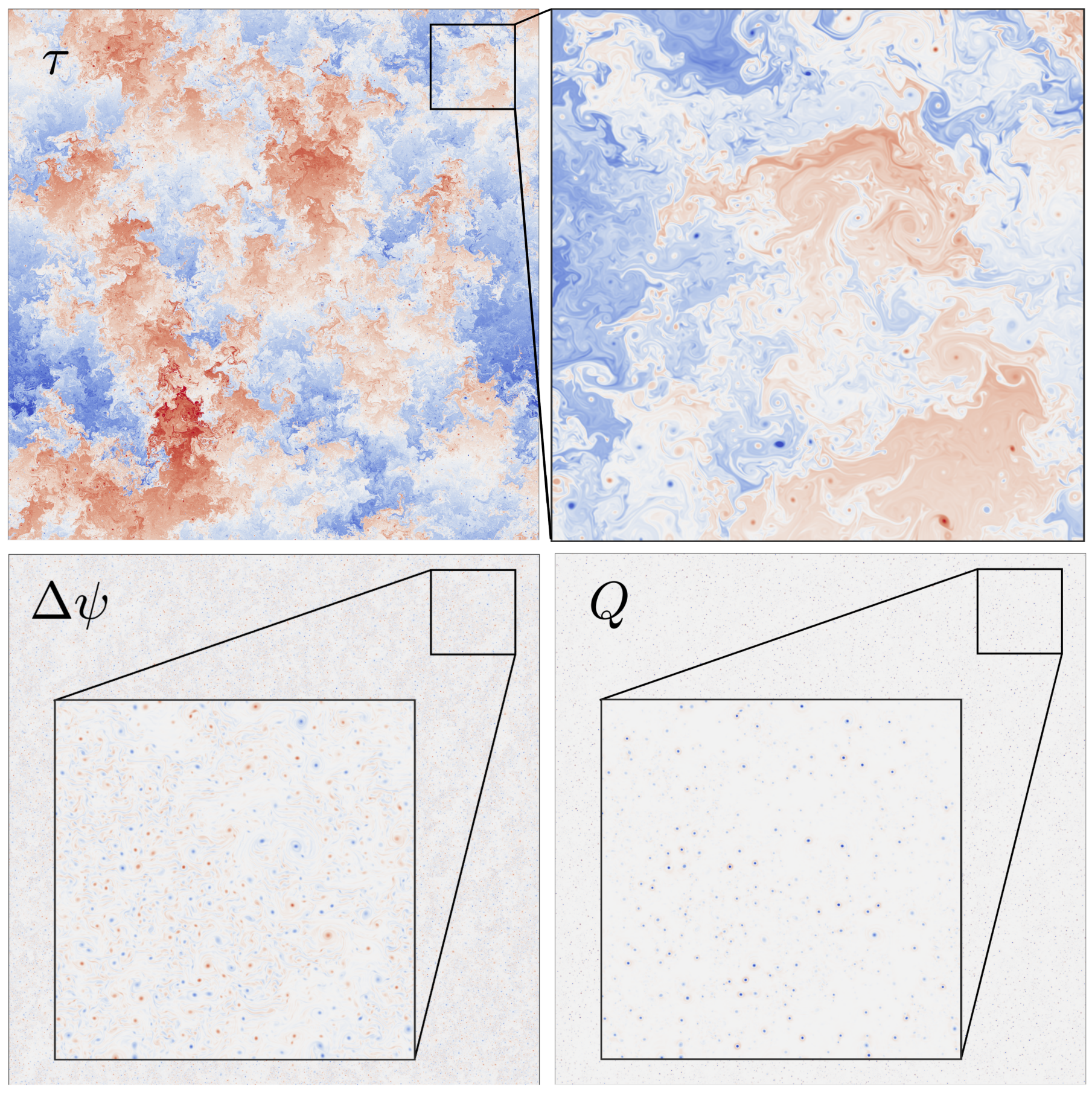}
\caption{\label{fig:snapshot} Snapshots of the passive tracer departure field $\tau$ (upper panels), of the vorticity field $\Delta \psi$ (bottom left) and of the Okubo-Weiss parameter $Q$  (bottom right) for a typical  $4096 \times 4096$ run with quadratic drag $\mu/k_f = 5 \times 10^{-4}$ and Kolmogorov forcing at $k_f  L = 400$ (positive values in red, negative values in blue, zero is white). \label{fig:snapshots}}\end{figure}

To diagnose the effective diffusivity of the flow, we consider the advection of a passive tracer subject to a uniform background gradient \cite{SMITH2002,grianik}. The tracer evolution equation being linear, without loss of generality we set the large-scale gradient to minus one and denote as $\tau(x,y,t)$ the doubly periodic departure from the background gradient. The evolution equation for $\tau$ then reads
\begin{eqnarray}
\partial_t \tau + J(\psi,\tau) = \psi_x - \nu_\tau \Delta^4 \tau  \, , \label{eq:tau}
\end{eqnarray}
where we employ the same hyperdiffusion coefficient for momentum and scalar concentration, $\nu_\tau=\nu$ (see Supplemental Material for a Table of all parameter values).  We perform DNS of equations (\ref{eq:zeta}) and (\ref{eq:tau}) using a pseudo-spectral solver running on GPU with resolution up to $8192 \times 8192$. Once the system has reached a statistically steady state, we extract $\epsilon$ together with the effective diffusivity of the flow. Because the background tracer gradient is set to $-1$, the effective diffusivity equals the tracer flux, $D=\langle \psi_x \tau \rangle$, where the angular brackets denote time and space average. The energy spectra, provided as End Matter, are very similar to the ones reported in previous studies~\cite{boffetta2000inverse,chen2006physical,boffetta2010evidence,boffetta2012two}, with best-fit exponents close to $-5/3$ for $k<k_f$. We focus on the large-domain small-hyperviscosity limit where $D$ is independent of both $L$ and $\nu$ (as assessed by running DNS with various $L$ and $\nu$). We illustrate this regime in Fig.~\ref{fig:snapshots}, where we show snapshots of the tracer departure field $\tau$, of the vorticity $\zeta$ and of the Okubo-Weiss parameter $Q=\psi_{xy}^2-\psi_{xx} \psi_{yy}$~\cite{okubo1970horizontal,weiss1991dynamics}. 

The inverse energy transfers are arrested by the drag term at a scale smaller than $L$. This prevents the formation of a large-scale condensate and rules out any `anomalous diffusion', as discussed in Ref.~\cite{biferale1995eddy}. The effective diffusivity $D$ depends only on the injection wavenumber $k_f$, the energy input rate $\epsilon$ and the drag coefficient $\kappa$ or $\mu$. In dimensionless form, we thus seek the dependence of the dimensionless diffusivity $\hat{D} = D \epsilon^{-1/3} k_f^{4/3}$ on the dimensionless drag coefficient $\hat{\kappa} =\kappa \, \epsilon^{-1/3} k_f^{-2/3}$ or $\hat{\mu} = \mu/ k_f$.

\textit{Diffusivity according to scale invariance.}— As discussed in Refs.~\cite{SMITH2002,grianik}, a simple dimensional estimate for the effective diffusivity can be deduced from the KLB assumption of a SIIC: as for any large-scale quantity, the effective diffusivity should depend on the energy flux $\epsilon$ and on the friction coefficient arresting the inverse energy cascade, while being independent of the small forcing scale $k_f^{-1}$.  Dimensional analysis then yields $D \sim \epsilon/\kappa^2$ for linear drag and $D\sim \epsilon^{1/3}/\mu^{4/3}$ for quadratic drag, which we recast in terms of the dimensionless quantities as~\cite{grianik}:
\begin{eqnarray}
 \hat{D} \sim  \hat{\kappa}^{-2} \, , \qquad \hat{D}  \sim  \hat{\mu}^{-4/3} \, , \label{eq:scalingKLB}
\end{eqnarray}
for linear and quadratic drag, respectively. 

\begin{figure*}
\includegraphics[scale = 0.45]{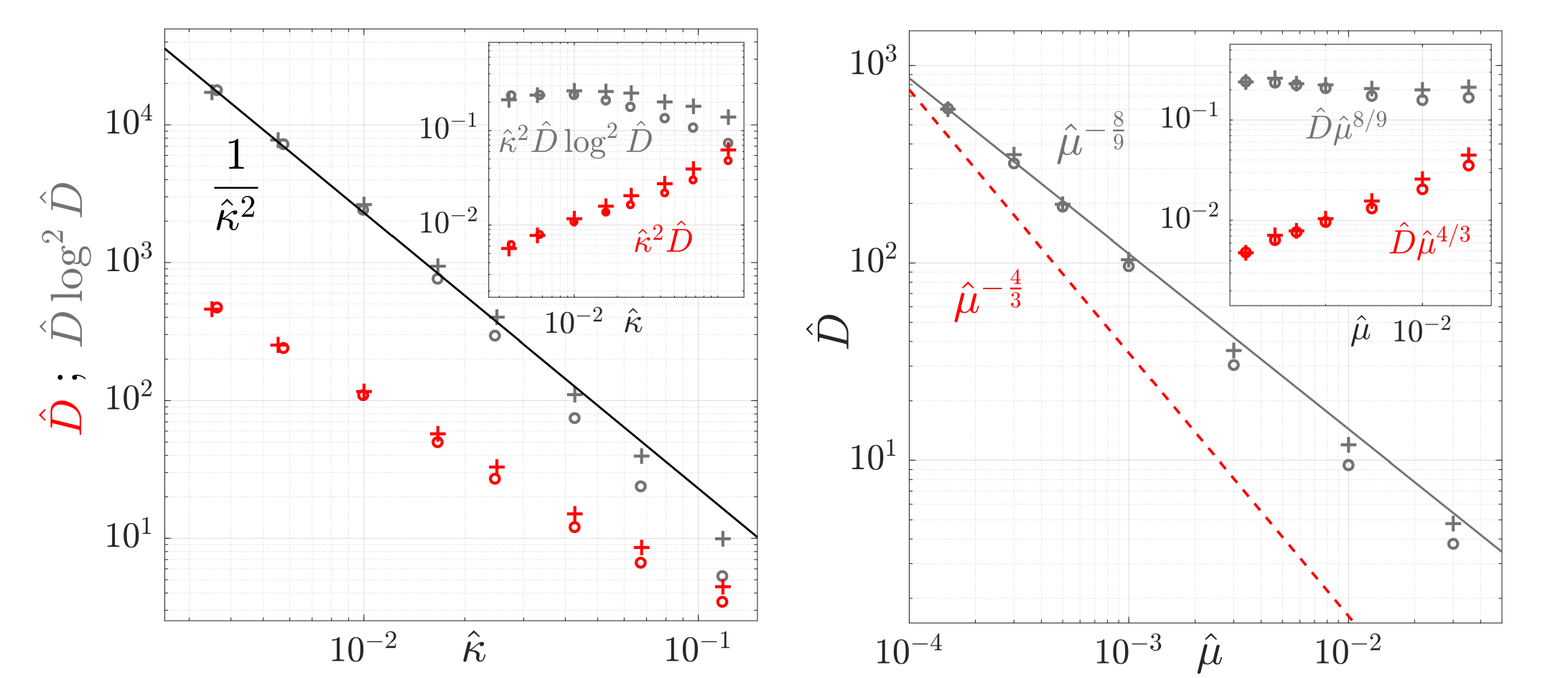}
\caption{\label{fig:scalingD} Dimensionless effective diffusivity $\hat{D}$ as a function of the dimensionless drag coefficient. Symbols are DNS data ($+$: white-noise forcing ; $\circ$: Kolmogorov forcing). In the left-hand panel we plot both $\hat{D}$ and $\hat{D} \log^2 \hat{D}$  as functions of $\hat{\kappa}$ to test the KLB prediction and the vortex-gas prediction, respectively. 
In the right-hand panel we plot the effective diffusivity as a function of the quadratic drag coefficient, together with the KLB prediction (dashed line) and the vortex-gas prediction (solid line). Inset: compensated plots using both the KLB prediction (red) and the present theory (gray).}
\end{figure*}

We compare these predictions to the DNS data in Fig.~\ref{fig:scalingD}. Suprisingly, the numerical data depart from the SIIC predictions (\ref{eq:scalingKLB}) for both linear and quadratic drag, as clearly illustrated by the compensated plots in the insets. In stark contrast with the assumption of complete scale-invariance, the diffusivity of the turbulent flow clearly retains some dependence on the small injection scale $k_f^{-1}$.

\textit{Diffusivity according to vortex-gas dynamics.—} A striking flow feature in physical space is the emergence of coherent vortices, as highlighted by the snapshot of $Q$ in Fig.~\ref{fig:snapshots}. With the goal of including them in a scaling theory, we model the coherent vortices using an idealized vortex-gas framework, in a similar fashion to GF~\cite{GalletFerrari,HadjerciGallet}. We consider a population of identical vortices of circulation $\pm \Gamma$ and core radius $r$, the typical vorticity within the vortex cores being $\zeta_\text{core}\sim \Gamma/r^2$. The vortex-gas is dilute, with a typical inter-vortex distance $\ell \gg r$. The vortices wander around as a result of mutual induction with a typical velocity $V \sim \Gamma / \ell$, the latter being also the typical large-scale velocity in the inter-vortex region between vortices. Assuming that the transport properties of a single dipole of oppositely signed vortices correctly reflect  transport within the vortex gas, GF show that:
\begin{equation}
D \sim \ell V \sim \Gamma \, . \label{eq:dipolartransport}
\end{equation}
In the inter-vortex region, energy injection and inverse transfers proceed following the standard KLB phenomenology~\cite{burgess2017vortex,wang2023coherent}. We thus estimate the large-scale velocity $V$ at the large inter-vortex distance~$\ell$ based on the KLB expression for the velocity increment at scale $\ell$:
\begin{equation}
V \sim (\epsilon \ell)^{1/3} \, . \label{eq:estimateV}
\end{equation}
Combining this relation with (\ref{eq:dipolartransport}) leads to expressions for $V$ and $\ell$ in terms of $D$ and $\epsilon$:
\begin{eqnarray}
V  \sim D^{1/4} \epsilon^{1/4} \, , & \qquad
\ell \sim D^{3/4} \epsilon^{-1/4} \, . \label{eq:vortexgas_lV}
\end{eqnarray}

The last two scaling arguments are based, respectively, on the energy and enstrophy power integrals. The spatially intermittent nature of frictional damping comes into play through the energy power integral. Hyperviscous energy dissipation being negligible, the energy power integral reads $\epsilon = \kappa \la {\bf u}^2 \ra$ for linear drag and $\epsilon = \mu \la |{\bf u}|^3 \ra$ for quadratic drag. Following GF, we estimate the second and third moments of the vortex-gas velocity field by considering a single isolated Rankine vortex of core radius $r$ and circulation $\Gamma$ located at the center of a disk-shaped domain of radius $\ell$. Computing the space average of ${\bf u}^2$ and $|{\bf u}|^3$ over the disk-shaped domain for this idealized velocity field yields $\la {\bf u}^2\ra \sim V^2 \log(\ell/r)$ and $\la |{\bf u}|^3\ra \sim V^3 \ell/r$, which we substitute into the energy power integral to obtain: 

\begin{equation}
\epsilon \sim \kappa V^2 \log \left( \frac{\ell}{r} \right) \, , \qquad \epsilon \sim   \mu V^3  \frac{\ell}{r}  \, , \label{eq:powerintegral}
\end{equation}
for linear and quadratic drag, respectively. Finally, an estimate for the core radius $r$ can be deduced from the enstrophy power integral. Following the vortex-gas description of freely evolving turbulence in Refs.~\cite{Carnavaleetal,weiss1999punctuated}, we adopt the picture of an isolated vortex being reinforced through continuous mergers with smaller intense vorticity structures~\cite{borue1994inverse}. This process leads to filamentation and enstrophy dissipation in the vicinity of the vortex core, a region of both strong vorticity and strong strain. The flux of enstrophy to small dissipative scales through this process is the product of the typical enstrophy in this region, $\zeta_\text{core}^2$, with the local strain rate $\Gamma/r^2 \sim \zeta_\text{core}$. We thus estimate the enstrophy flux as $\zeta_\text{core}^3$ within the near-core regions, that is, over a fraction $r^2/\ell^2$ of the domain. The contribution of the near-core regions to the space-averaged enstrophy dissipation rate is thus estimated as $\zeta_\text{core}^3 \, r^2/\ell^2$. Demanding that the latter be less than or equal to the enstrophy injection rate $k_f^2 \epsilon$ leads to the following inequality for the vortex core radius:
\begin{eqnarray}
r \lesssim \frac{k_f \epsilon^{1/2} \ell}{\zeta_\text{core}^{3/2}} \, . \label{eq:ineqr}
\end{eqnarray}
As discussed in Ref.~\cite{Carnavaleetal}, isolated vortices within the vortex gas reinforce and expand through mergers, conserving energy and core vorticity, while dissipating enstrophy. They can do so as long as enstrophy dissipation by the mergers is smaller than enstrophy injection by the forcing. In the equilibrated state, we thus expect inequality (\ref{eq:ineqr}) to be saturated. We thus replace the inequality (\ref{eq:ineqr}) with an equality, before combining the resulting relation with (\ref{eq:dipolartransport}) and (\ref{eq:vortexgas_lV}), together with $\zeta_\text{core} \sim \Gamma/r^2 \sim D/r^2$. This leads to the following estimate for the dimensionless core radius:
\begin{eqnarray}
k_f r \sim \hat{D}^{3/8} \, . \label{eq:rcore}
\end{eqnarray}

Starting from the energy power integral (\ref{eq:powerintegral}), one finally substitutes the estimates (\ref{eq:vortexgas_lV}) and (\ref{eq:rcore}) for $V$, $\ell$ and $r$. This leads to the following scaling predictions for the eddy diffusivity of the flow in terms of the dimensionless drag coefficient:
\begin{eqnarray}
 \hat{D} \log^2 \hat{D}  \sim  \hat{\kappa}^{-2} \, , \qquad \hat{D}  \sim  \hat{\mu}^{-8/9} \, , \label{eq:scalingVG}
\end{eqnarray}
for linear and quadratic drag, respectively.
In contrast with the assumption of scale invariance, reverting to dimensional quantities indicates that the predicted eddy diffusivity (\ref{eq:scalingVG}) explicitly involves the forcing scale: $D \log^2[ D \epsilon^{-1/3} k_f^{4/3}]\sim \epsilon/\kappa^2$  or $D \sim \epsilon^{1/3} k_f^{-4/9} \mu^{-8/9}$. As shown in Fig.~\ref{fig:scalingD}, the predictions~(\ref{eq:scalingVG}) capture the numerical data with excellent accuracy in the low-drag regime, for both types of drag and both forcing protocols.\\

\textit{Discussion: connection to baroclinic turbulence.—} Interestingly, the present scaling theory sheds new light on traditional approaches to parameterize `baroclinic' turbulence in oceans and atmospheres~\cite{held1999macroturbulence,LarichevHeld,SMITH2002,ChangHeld}. Baroclinic turbulence is a form of quasi-2D turbulence driven by an instability mechanism that injects kinetic energy around a small spatial scale $\lambda$ (the Rossby deformation radius~\cite{salmon1998lectures,pedlosky2013geophysical,vallis2017atmospheric,gallet2022transport}). The 2D velocity field is coupled to a temperature field $\tau$, which has dimensions of a streamfunction. The latter is subject to a background meridional gradient denoted as $U$ in this context. The source of kinetic energy in baroclinic turbulence is the meridional transport of heat, hence a direct proportionality relation between the mean meridional heat flux and the mean kinetic energy dissipation rate of the flow, $\epsilon = D U^2 / \lambda^2$. The sink of kinetic energy is linear or quadratic drag. The dimensional control parameters are $U$, $\lambda$ and the drag coefficient in this context, and scaling theories aim at relating the dimensionless diffusivity $D_*=D/(U \lambda)$ to the dimensionless drag coefficient $\kappa_*=\kappa \lambda/U$ or $\mu_*=\mu \lambda$. Although the equations governing baroclinic turbulence depart from the standard 2D Navier-Stokes equation, the traditional approach to deriving a scaling theory is based on a parallel between the two systems: one assumes that the forcing wavenumber is $k_f \sim \lambda^{-1}$ and combines the relation $\epsilon = D U^2 / \lambda^2$ with estimates for $D$ based on the phenomenology of 2D turbulence. However, combining this relation with the KLB prediction (\ref{eq:scalingKLB}) for quadratic drag leads to $D_*\sim 1/\mu_*^2$, a prediction clearly invalidated by DNS of baroclinic turbulence~\cite{ChangHeld,GalletFerrari,HadjerciGallet,hadjerci2024two} (a similar failure arises for linear drag). In parallel, GF and later Hadjerci \& Gallet (\cite{HadjerciGallet}, HG in the following) designed a purely vortex-based theory that captures the behavior of $D_*$ with drag $\kappa_*$ or $\mu_*$ with excellent accuracy. The puzzle remained why baroclinic turbulence would behave so differently from standard 2D Navier-Stokes turbulence. The present study solves this puzzle by insisting that emergent coherent vortices are, in fact, a crucial ingredient of standard 2D Navier-Stokes turbulence too. When the relations $\epsilon = D U^2 / \lambda^2$ and $k_f \sim \lambda^{-1}$ are combined with the new scaling predictions (\ref{eq:scalingVG}) for the 2D Navier-Stokes diffusivity, one recovers the successful predictions of HG for the dependence of $D_*$ on $\kappa_*$ or $\mu_*$.\\

\textit{Conclusion.—} We have reported a suite of DNS of 2D Navier-Stokes turbulence with drag, focussing of the effective diffusivity of the flow. By reaching lower drag than previous studies~\cite{SMITH2002,grianik}, we identified clear departures from the SIIC dimensional estimates. We introduced an alternate scaling theory that takes into account the coherent vortices that populate 2D turbulent flows. The vortices derail the SIIC scaling predictions because they induce localized frictional dissipation within the small vortex cores. The combination of broken scale-invariance and spatially intermittent energy dissipation is somewhat reminiscent of intermittency effects in the forward energy cascade of 3D turbulence. This parallel shall be drawn with care, however, as velocity structure functions are known to display no intermittency corrections in the inertial range of the 2D inverse energy cascade~\cite{boffetta2000inverse}.

\appendix

\section{End matter: Energy spectra}

\begin{figure}[h!]
\includegraphics[width=8.5cm]{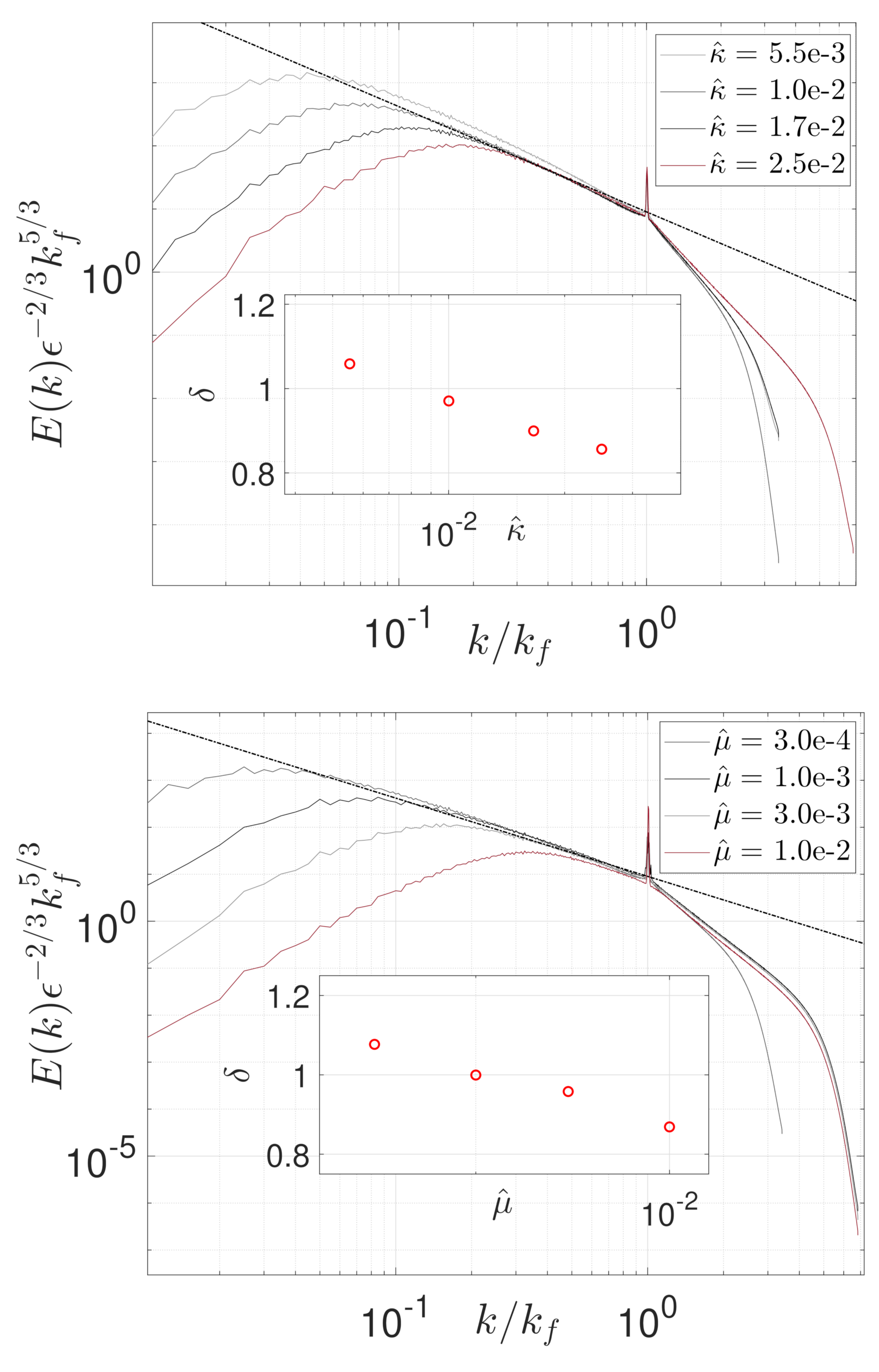}
\caption{Energy spectra from sample simulations with linear (top) and quadratic (bottom) drag, driven by white-noise-in-time forcing. The spectra are non-dimensionalized using $\epsilon$ and $k_f$. The inset displays the ratio $\delta$ of the best-fit exponent $\alpha$ to the KLB prediction $-5/3$.}
\label{fig:spectra}\end{figure}

In Fig.\ref{fig:spectra} we plot the energy spectra extracted from DNS performed with white-noise forcing and both types of drag, together with a $k^{-5/3}$ eyeguide. As expected, the spectra are very similar to the ones reported in previous numerical studies with white-noise forcing and (linear) drag, see e.g. Ref.~\cite{boffetta2010evidence}. To extract a spectral slope in the inverse cascade range, we first introduce an energy-containing wavenumber $k_L^2=\la {\bf u}^2 \ra/\la \psi^2 \ra$, before performing a power-law fit to each spectrum over the range $k \in [k_L, k_f]$. The resulting power-law exponent $\alpha$ is compared to the KLB value $-5/3$ in the inset of each panel, where we plot the ratio $\delta=\alpha/(-5/3)$. The best-fit exponent $\alpha$ is within 15\% percents of the theoretical value $-5/3$ at low drag.

\bibliographystyle{unsrt}
\bibliography{main}

\end{document}